\documentclass[aps,pre,twocolumn,letterpaper]{revtex4}

\usepackage{amsmath}

\usepackage{graphicx}
\usepackage{indentfirst}

\newcommand{\sFrac}[2]{{\textstyle\frac{#1}{#2}}}

\newcommand{\calBold}[1]{\mbox{\boldmath${\cal #1}$}}

\newcommand{\ket}[1]{|#1\rangle}
\newcommand{\bra}[1]{\langle #1|}
\newcommand{\braket}[2]{\langle #1|#2\rangle}

\begin{document}
\title{Simulations of driven overdamped frictionless hard spheres}
\author{Edan Lerner, Gustavo D{\"u}ring, and Matthieu Wyart}
\affiliation{Center for Soft Matter Research, Department of Physics, New York University, New York, NY 10003}
\begin{abstract}
We introduce an event-driven simulation scheme for overdamped dynamics of frictionless hard spheres
subjected to external forces, neglecting hydrodynamic interactions. Our event-driven approach is based
on an exact equation of motion which relates the driving force to the resulting velocities through the
geometric information characterizing the underlying network of contacts between the hard spheres. Our
method allows for a robust extraction of the instantaneous coordination of the particles as well as contact
force statistics and dynamics, under any chosen driving force, in addition to shear flow and compression.
It can also be used for generating high-precision jammed packings under shear, compression, or both. We
present a number of additional applications of our method.
\end{abstract}
\maketitle

\section{Introduction}
Systems of frictionless hard spheres serve as prototypical models in statistical physics, displaying a variety of emergent complex phenomena \cite{86HM,62AW,00TS,00CR,12IBS,07BW}. Owing their popularity to their inherent simplicity, systems of frictionless hard spheres are used to model gases \cite{82Leu}, supercooled liquids \cite{07BW}, dense suspensions  \cite{07OT,11OT,09HB,12LDW} and granular media \cite{00CR,00Roux}. The simplicity of hard-sphere systems stems from the absence of energy scales in the hard-sphere interactions. It is this simplicity which makes these systems a preferred choice of model for the investigation of complex phenomena in particulate systems.

Perhaps the most extensively studied model of hard spheres is the fully inertial fluid, in which collisions between particles are entirely elastic. In these systems, collisions are instantaneous, i.e., colliding particles spend no time at all in contact, but instead conservation of energy and momentum determines the postcollision velocities as a function of the pre-collision velocities. Event-driven simulations of these systems are carried out by predicting the next collision time from the instantaneous velocities and positions of the particles \cite{91AT}. Then, the system is evolved forward in time directly to the next collision. 

Here we shall rather focus on non-Brownian hard particles immersed in a viscous fluid, in the overdamped limit where
inertia is negligible. We shall assume further that hydrodynamic interactions are negligible. This assumption has been made in the
context of flow near jamming \cite{12IBS,07OT,11OT,09HB,12LDW}, where it appears to capture at least qualitatively the critical behavior in the dense limit \cite{12LDW}. Within the framework of these assumptions, contacts are formed between the particles upon collisions, as they are pushed towards each other. These contacts persist for finite intervals of time, during which repulsive forces are exerted between the particles in contact.

Our simulational approach is based on the exactly derivable equations of motion of overdamped dynamics in the hard-sphere
limit, neglecting hydrodynamic interactions. These equations are used to build an event-driven simulation. The equations of motion are entirely based on geometric information, which allows us to calculate the contact forces between the constituent hard particles, 
and hence their distribution and evolution \cite{12LDWb}. The main idea is to evolve the system according to the equations of motion, while carefully handling the formation of new contacts and the opening of existing contacts between particles. Our method shares some similarities with the method of contact dynamics \cite{09RR} used in dry granular materials, in which contact forces and velocities are resolved iteratively under a set of complementarity relations. 

The simulational scheme presented below has merits and disadvantages with respect to existing methods. In terms of bare complexity, the method we present is far inferior to conventional
molecular dynamics methods, in which the running time typically scales linearly with the number of particles, given that interactions
are sufficiently short ranged. The running time of the scheme presented here scales at least quadratically with the number of 
particles. This is a consequence of the event-driven nature of the scheme, together with the effectively long-range interactions
which can span the entire system at high packing fractions. However, when compared to existing methods in which hardsphere dynamics is approached by reducing the loading rates in systems of soft-potential interactions \cite{07OT,11OT,09HB}, our method has the advantage of directly sampling the hard-sphere limit. Furthermore, our method requires essentially no adjustable parameters; particle trajectories are invariant when length and time are measured in the appropriate microscopic units.

The formalism presented here allows for useful extensions of the methods built in this work. For instance, it is straightforward
to introduce hard boundaries (such as various container shapes, hoppers, or inclined planes), or constructing composite particles
made of any number of beads, which can be floppy or constrained as rigid bodies. The method can also be used to generate jammed
packings under various loading geometries, while maintaining complete control over the identity of particles in contact, in
distinction from existing methods  \cite{91LSP,03OSLN}. This control allows us to
properly account for rattlers in jammed states, with no ambiguity whatsoever.

This paper is organized as follows. Section~\ref{equationOfMotion} derives the equations of motion for overdamped, frictionless hard spheres, either driven by external forces, or under imposed spatial deformations, neglecting hydrodynamic interactions. We also include an extension of our formalism for simple shear flow under fixed imposed pressure. Section~\ref{contactNetwork} illustrates the concept of the instantaneous contact network, which is geometric information on which the equation of motion is based. Section~\ref{simulationDetails} contains an elaborate description of our simulational scheme and ends with a prescription for generating jammed configurations under shear or compression. Section~\ref{results} describes applications of our simulation method and presents some results from our simulations, illustrating the utility of our event-driven approach. 
Section~\ref{conclusions} presents concluding remarks.

\section{Equations of motion for overdamped, frictionless, driven hard spheres}
\label{equationOfMotion}
Consider a system of $N$ frictionless hard spheres (referred to in the following as particles)
in a volume $\Omega$ with periodic boundary conditions in $d$ dimensions, such that there 
are no overlapping particles, and 
some of the particles are exactly in contact: the distance between their
centers is equal to the sum of their radii. When the system is unjammed, 
the number of contacts $N_c$ in the system
remains smaller than the number of spatial degrees of freedom $Nd-d$. Note that we 
subtract $d$ translations but not rotations due to the periodic boundary conditions.
We denote the $d$ dimensional vector
of the $i$'th particle coordinates as $\vec{R}_i$, its time derivative
as $\vec{V}_i$, and define the 
directional differences $\vec{R}_{ij} = \vec{R}_j - \vec{R}_i$, 
the pairwise distances $r_{ij} = \sqrt{\vec{R}_{ij}\cdot\vec{R}_{ij}}$, and
the normalized directions $\vec{n}_{ij} = \vec{R}_{ij}/r_{ij}$. 
We will refer to the \emph{contact network} as the set of all pairs of particles that are in contact
at some instance in time, and the geometric information that accompanies the network, 
namely the directions $\vec{n}_{ij}$, and the pairwise distances $r_{ij}$.

We begin the derivation with accounting for the hard-sphere interactions; given some vector of particles' velocities~$\vec{V}_k$, the rate of change induced on a pairwise distance $r_{ij}$ is
\begin{equation}\label{foo01}
\dot{r}_{ij} = \!\sum_k\frac{\partial r_{ij}}{\partial\vec{R}_k}\cdot \vec{V}_k = 
\!\sum_k(\delta_{jk} - \delta_{ik})\,\vec{n}_{ij}\cdot \vec{V}_k =
(\vec{V}_j - \vec{V}_i) \cdot \vec{n}_{ij} \ .
\end{equation}
The above relation consists of a linear transformation of vectors from the space of the particles (of dimension $Nd$) to vectors in the
space of contacts (of dimension $Nc$). We define 
\begin{equation}\label{esDefinition}
{\cal S} \equiv \frac{\partial r_{ij}}{\partial\vec{R}_k}\ ;
\end{equation}
then Eq.~(\ref{foo01}) can be written in bra-ket notation as 
\begin{equation}
\ket{\dot{r}} = {\cal S}\ket{V}\ .
\end{equation}
Here and in the following we use bra-ket notations with uppercase letters to denote vectors from the space of the particles (e.g.,
$\ket{V}$ for particle velocities), and bra-ket notations with lower-case 
letters to denote vectors from the space of contacts (e.g., $\ket{r}$ for the vector of pairwise distances). 

As the particles are completely rigid, they cannot penetrate each other. We thus impose a constraint on the particles' velocities:
they must keep the distance between the pairs of particles that are in contact unchanged, if the contact force exerted between the
pairs of particles is positive. We will see in Section 3 how a given set of contacts between the particles does not change except at
discrete points in time. Except for at those discrete time points, the
velocities must satisfy
\begin{equation}\label{constraint}
\ket{\dot{r}} = {\cal S}\ket{V} = 0\ .
\end{equation}
In the next two subsections we present the derivation of the
equations of motion for the case of dynamics under external
forces, and the case of spatial deformations of the system (e.g.,
compression or shear). Section \ref{constantPressure} contains an extension of our
formalism to systems of hard particles under simple shear, with
fixed imposed pressure, as opposed to fixed volume.

\subsection{Overdamped dynamics under external forces}
In our framework, there are three forces acting upon each particle: a drag force $\vec{F}_k^{\rm drag}$, 
the force exerted on a particle by its neighbors with which it is in contact $\vec{F}_k^{\rm cont}$,
and some external driving force $\vec{F}_k^{\rm ext}$. Assuming that the dynamics is overdamped,
and neglecting hydrodynamic interactions, the net force on each particle must always be zero:
\begin{equation}\label{forceBalance1}
\vec{F}_k^{\rm ext} + \vec{F}_k^{\rm drag} + \vec{F}_k^{\rm cont} = 0\ .
\end{equation}
We assume conventional Stokes drag forces acting upon particles,
which are opposite in sign and proportional to their velocities:
\begin{equation}\label{dragForces}
\vec{F}_k^{\rm drag} = -\xi_0^{-1}\vec{V}_k\ ,
\end{equation}
where $\xi_0$ has units of $\sFrac{\rm time}{\rm mass}$. $\xi_0$ may generally depend on the
radius of the $k$'th particle; however, for the sake of brevity, we consider here monodisperse spheres, as the extension to polydisperse spheres is straightforward. Denoting the magnitude of the
(purely repulsive) force exerted between the $j$'th and $k$'th particles as $f_{jk}(=f_{kj})$, the forces exerted on a particle by its neighbors with
which it is in contact can be written as
\begin{eqnarray}
\vec{F}_k^{\rm cont} & = & \sum_{j {\mbox{\tiny{ in contact with }}} k}\vec{n}_{jk}f_{jk}  \nonumber \\
& =  & \sum_{\mbox{\tiny{all pairs $i,j$ in contact}}}(\delta_{jk} - \delta_{ik})\,\vec{n}_{ij}f_{ij}  \nonumber \\ 
& =  & \sum_{\mbox{\tiny{all pairs $i,j$ in contact}}}\frac{\partial r_{ij}}{\partial \vec{R}_k}f_{ij}\ , \label{contactForces1}
\end{eqnarray}
The above equation, similarly to Eq.~(\ref{foo01}), also consists of a linear
transformation, but this time from the space of contacts to the
space of particles, with the \emph{transpose} of the same linear operator ${\cal S}$ of Eq.~(\ref{esDefinition}); we thus write Eq.~(\ref{contactForces1}) as
\cite{00Roux,05Wyart}
\begin{equation}\label{moo00}
\ket{F^{\rm cont}} = {\cal S}^T\ket{f} \ ,
\end{equation}
with $\ket{f}$ a vector of dimension $N_c$ denoting the contacts forces $f_{ij}$.
Inserting Eqs.~(\ref{dragForces}) and (\ref{contactForces1}) in Eq.~(\ref{forceBalance1}) and rearranging, we find
\begin{equation}\label{velocities2}
\ket{V} = \xi_0\ket{F^{\rm ext}} + \xi_0{\cal S}^T\ket{f}\ .
\end{equation}
Eq.~(\ref{velocities2}) relates the velocities $\ket{V}$ to the contact forces $\ket{f}$ and the external force 
$\ket{F^{\rm ext}}$. In the absence of contacts, particles will simply move in the 
direction of the external force $\ket{F^{\rm ext}}$. 
We next follow Eq.~(\ref{constraint}) and operate on Eq.~(\ref{velocities2}) with ${\cal S}$, requiring
that this operation vanishes:
\begin{equation}\label{foo02}
{\cal S}\ket{V} = \xi_0{\cal S}\ket{F^{\rm ext}} + \xi_0{\cal S}{\cal S}^T\ket{f} = 0\ .
\end{equation}
The operator ${\cal S}{\cal S}^T \equiv {\cal N}$ is symmetric and semi-positive definite, hence
unless it is singular (see Subsect.~\ref{jammed} for discussion), it is invertible.
We can thus invert Eq.~(\ref{foo02}) in favor of the contact forces:
\begin{equation}\label{contactForces}
\ket{f} = -{\cal N}^{-1}{\cal S}\ket{F^{\rm ext}}\ .
\end{equation}
With the above explicit expression for the contact forces, Eq.~(\ref{velocities2}) provides us with
the velocities
\begin{eqnarray}\label{eom}
\ket{V} & = & \xi_0\ket{F^{\rm ext}} - \xi_0{\cal S}^T{\cal N}^{-1}{\cal S}\ket{F^{\rm ext}} \nonumber \\
& = & \xi_0\left( {\cal I} - {\cal S}^T{\cal N}^{-1}{\cal S}\right)\ket{F^{\rm ext}} \ ,
\end{eqnarray}
with ${\cal I}$ being the identity operator. 
This concludes the derivation of the equations of motion for the case of external driving forces; the contact forces
and the instantaneous velocities are completely determined 
by the external driving force $\ket{F^{\rm ext}}$ and by the geometry 
of the underlying contact network encoded in the operator~${\cal S}$.

\subsection{Overdamped dynamics under imposed spatial deformations}
\label{spatialDeformations}
The equation of motion (\ref{eom}) provides us with the instantaneous velocities 
for our system subjected to an external driving force $\ket{F^{\rm ext}}$. 
If instead one wishes to impose some mode of strain on the system,
e.g.~shear or compression, the equations must be slightly modified to account for the 
imposed deformation instead of the driving forces. The following subsection 
derives the equations of motion for frictionless hard spheres under spatial deformations,
assuming that the particles are immersed in a \emph{flowing viscous fluid}, 
described by an affine velocity field $\ket{V^{\rm aff}}$. Hydrodynamic interactions are neglected.
This simple model has been originally proposed by Durian
\cite{95Dur,97Dur}, and recently popularized by Olsson and Tietel \cite{07OT,11OT} and others \cite{09HB,12LDW,09Hat} as a model 
for non-Brownian suspension flows. The framework presented in the previous subsection naturally allows for the 
extension to spatial deformations within this model. 

The strain rate tensor $\dot{\calBold{\epsilon}}$ is related to the affine velocity field via
\begin{equation}\label{moo03}
\left(\vec{V}^{\rm aff}_j - \vec{V}^{\rm aff}_i\right)\cdot \vec{n}_{ij} = \frac{\vec{R}_{ij}\cdot\dot{\calBold{\epsilon}}\cdot\vec{R}_{ij}}{r_{ij}}\ .
\end{equation}
The affine velocity field $\ket{V^{\rm aff}}$ should be chosen such that the desired deformation geometry is acheived. 
For instance, for simple shear in the $x,y$-plane one can choose $\vec{V}^{\rm aff} = \dot{\gamma}y\hat{x}$,
with $\dot{\gamma}$ the strain-rate. For compression, one can choose $\vec{V}^{\rm aff} = -\dot{\gamma}\vec{R}$. 
Special care should be taken when imposing periodic boundary conditions, since the velocities $\ket{V^{\rm aff}}$ must 
be continuous across the boundaries. The operation ${\cal S}\ket{V^{\rm aff}}$
results in a contact-space vector which contains the rate of change of pairs of particles in contact due
to the imposed deformation; contacts which pass across the boundary should be accounted for properly.
In the case of simple shear, continuity of the velocities can be assured by employing Lees-Edwards boundary conditions~\cite{91AT}.

The drag forces are now considered to be proportional to the {\bf difference} between 
the imposed affine velocities and the velocities of the particles \cite{07OT,11OT,09HB,95Dur,97Dur,09Hat,12LDW}, i.e.
\begin{equation}\label{strainDragForces}
\ket{F^{\rm drag}} = - \xi_0^{-1}\left(\ket{V} - \ket{V^{\rm aff}}\right)\ .
\end{equation}
Using Eqs.~(\ref{moo00}) and (\ref{strainDragForces}), the force-balance equation now reads
\begin{equation}
\ket{F^{\rm cont}} + \ket{F^{\rm drag}} = {\cal S}^T \ket{f} - \xi_0^{-1}\left(\ket{V} - \ket{V^{\rm aff}}\right) = 0\ .
\end{equation}
From here the velocities are extracted as
\begin{equation}\label{moo01}
\ket{V} = \ket{V^{\rm aff}} + \xi_0{\cal S}^T \ket{f}\ .
\end{equation}
To account for the hard-particle interactions, we again follow Eq.~(\ref{constraint}) and require that 
the velocities $\ket{V}$ keep the distances between pairs of particles in contact unchanged, namely 
\begin{equation}
\ket{\dot{r}} = {\cal S}\ket{V} = {\cal S}\ket{V^{\rm aff}}  + \xi_0{\cal S}{\cal S}^T \ket{f} =
{\cal S}\ket{V^{\rm aff}}  + \xi_0{\cal N}\ket{f} = 0\ .
\end{equation}
Inverting in favor of the contact forces $\ket{f}$
\begin{equation}\label{strainContactForces}
\ket{f} = -\xi_0^{-1}{\cal N}^{-1}{\cal S}\ket{V^{\rm aff}}\ .
\end{equation}
Inserting this relation in Eq.~(\ref{moo01}), we obtain an expression for the velocities
\begin{equation}\label{strainVelocities}
\ket{V} = \left( {\cal I} - {\cal S}^T{\cal N}^{-1}{\cal S}\right)\ket{V^{\rm aff}}\ . 
\end{equation}

\begin{figure*}[ht]
\centering
\includegraphics[scale = 0.50]{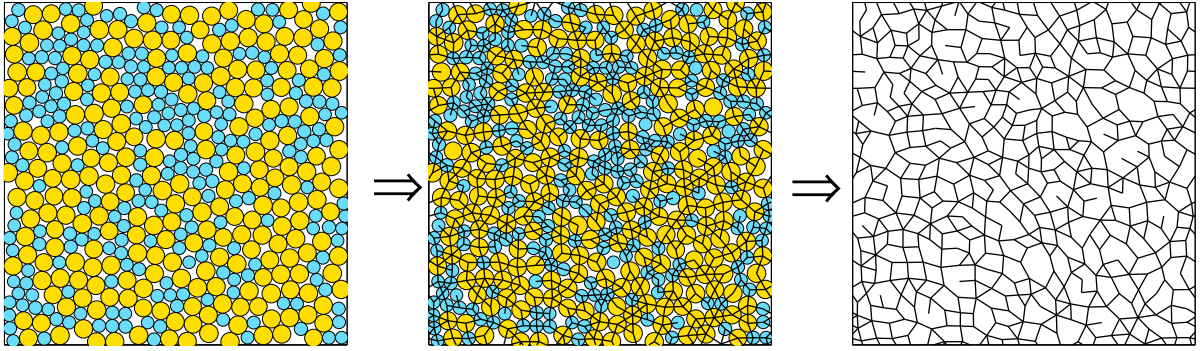}
\caption{\footnotesize The instantaneous contact network is defined at any instance in time
by collecting the set of all pairs of particles in contact, and extracting 
the geometric information of the corresponding network.}
\label{contactNetworkFig}
\end{figure*}

\subsection{Constant-pressure shear-flow}
\label{constantPressure}
Our formalism allows for a simple extension of the spatial deformations framework of the previous Subsection
for simple shear, such that the imposed pressure can be fixed, while allowing for volume fluctuations,
much like in molecular dynamics simulations of thermal systems under fixed pressure~\cite{91AT}. 
The idea is to add a compressional component to the deformation rate tensor such that the resulting 
pressure under simple shear will be set at some pre-determined fixed pressure~$p$. 
To this aim, we characterize the affine flow by the deformation rate tensor $\dot{\calBold{\epsilon}}$
parametrized (in two-dimensions, with a trivial extension to three-dimensions) as 
\begin{equation}\label{moo02}
\dot{\calBold{\epsilon}} \to \left(
\begin{array}{cc}
-\kappa \dot{\gamma} & \dot{\gamma}/2 \\
\dot{\gamma}/2 & -\kappa \dot{\gamma} 
\end{array}\right) \ .
\end{equation}
The parameter $\kappa$ stands for the ratio between the rate of isotropic 
compression and the rate of simple shear. A suitable choice for the affine velocities is
\begin{equation}\label{constantPressureAffineVelocities}
\vec{V}^{\rm aff} = \dot{\gamma}\left(R_y\hat{x} + \kappa\vec{R}\right)\ .
\end{equation}
As stated in Subsection~(\ref{spatialDeformations}), the above affine velocity field should be 
understood to include the necessary modifications of the size and shape of the box, and the associated changes in the
periodic boundary conditions \cite{91AT}.  
It is easily verified that Eq.~(\ref{moo03}) is satisfied with the above choices for  $\dot{\calBold{\epsilon}}$
and $\vec{V}^{\rm aff}$; the operation of ${\cal S}$ on the affine velocities is
\begin{equation}
{\cal S}\ket{V^{\rm aff}} = -\kappa\dot{\gamma} \ket{r} +  \dot{\gamma}\ket{\sFrac{r_xr_y}{r}}\ ,
\end{equation}
where $\ket{\sFrac{r_xr_y}{r}}$ is a vector in contact-space of $N_c$ components, with each
entry describing how the particular contact changes under simple shear in the $x,y$-plane, namely 
$\frac{(\vec{R}_{ij}\cdot\hat{x})(\vec{R}_{ij}\cdot\hat{y})}{r_{ij}}$. Following Eq.~(\ref{strainContactForces}), 
the contact forces are given by 
\begin{equation}\label{constantPressureForces}
\ket{f} = \dot{\gamma}\xi_0^{-1}\left( \kappa{\cal N}^{-1}\ket{r} + {\cal N}^{-1}\ket{\sFrac{r_xr_y}{r}} \right)\ .
\end{equation}
The pressure is thus given by \cite{91AT}
\begin{eqnarray}
p & = & \frac{1}{\Omega d}\sum_{{\mbox{\tiny{contacts }}}ij}
f_{ij}r_{ij} = \frac{\braket{f}{r}}{\Omega d} \nonumber \\
&= &\frac{\dot{\gamma}}{\xi_0\Omega d}\left(\kappa\bra{r}{\cal N}^{-1}\ket{r} - \bra{\sFrac{r_xr_y}{r}}{\cal N}^{-1}\ket{r}\right)\ .
\end{eqnarray}
Solving for $\kappa$:
\begin{equation}\label{kappa}
\kappa = \frac{p\,\xi_0\dot{\gamma}^{-1}\Omega d + \bra{r}{\cal N}^{-1}\ket{\sFrac{r_xr_y}{r}}}{\bra{r}{\cal N}^{-1}\ket{r}}\ .
\end{equation}
This fully determines the instantaneous deformation-rate tensor $\dot{\calBold{\epsilon}}$, and the associated
affine velocities given in Eq.~(\ref{constantPressureAffineVelocities}). The velocities are still given by 
Eq.~(\ref{strainVelocities}), with the affine velocities (which depend on $\kappa$) as given by
Eq.~(\ref{constantPressureAffineVelocities}).

\subsection{Invariance of particle trajectories}
As mentioned in the introduction, the overdamped dynamics we consider here results in invariance of particle-trajectories 
under changes of the magnitude of the driving force $f^{\rm ext} \equiv \sqrt{\braket{F^{\rm ext}}{F^{\rm ext}}}/N$ 
or the parameter $\xi_0$, if time is measured in units of $\tau \equiv \lambda/(f^{\rm ext}\xi_0)$, with $\lambda$ the particles' diameter. 
In the case of spatial deformations, measuring time in units of $\tau \equiv \dot{\gamma}^{-1}$
and forces in units of $\lambda\dot{\gamma}/\xi_0$ keeps the physics invarient to changes in $\dot{\gamma}$
or $\xi_0$.

\section{Instantaneous contact network}
\label{contactNetwork}
\begin{figure}[!ht]
\centering
\includegraphics[scale = 0.35]{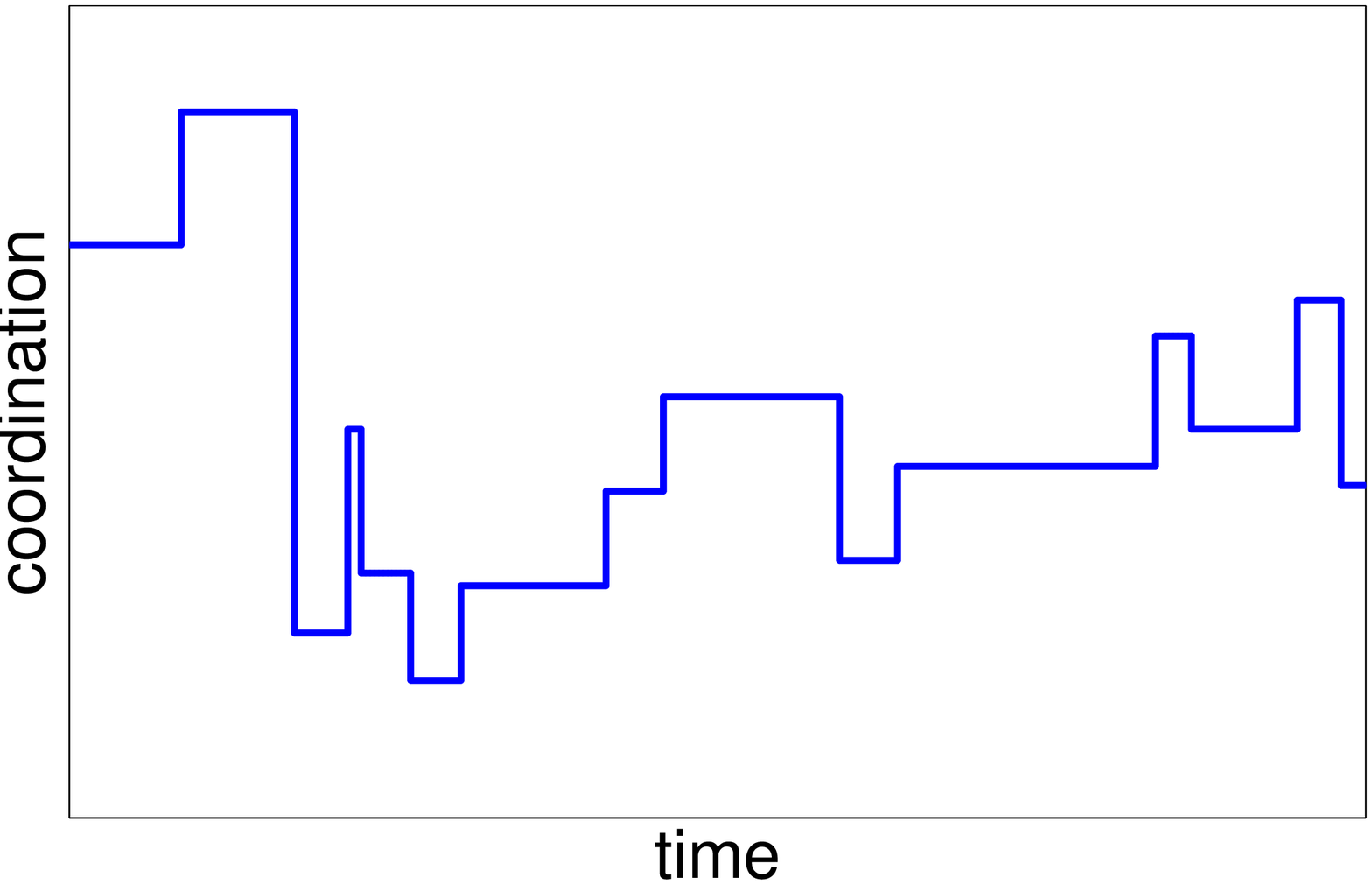}
\caption{\footnotesize A typical signal of the instantaneous average coordination evolving in time. 
A contact network remains intact during the time intervals between events of contact formation/breaking.}
\label{coordination}
\end{figure}

The equations of motion derived above hold for a single instance in time.
They are entirely based on the geometric information encoded in the operator ${\cal S}$,
which is defined on the existing set of contacts at that time, and on the driving force $\ket{F^{\rm ext}}$
or the imposed spatial deformation encoded in $\ket{V^{\rm aff}}$. 
Given a configuration of the system we consider all the pair of particles in contact
as our contact network (see Fig.~\ref{contactNetworkFig}), and based on this set we calculate the operator ${\cal S}$.
However, this set of contacts is ever-changing, as particles move apart
or collide. Nevertheless, the events of `breaking' a contact between two particles, 
or the formation of a new contact between two colliding particles, are instantaneous.
As a result, one can define time intervals in which a specific contact network
remains intact. These time intervals are exactly the periods between 
any such event of contact-breaking or contact-formation, as illustrated in Fig.~\ref{coordination}. 

The idea of an instantaneous contact network is central to our simulation method. 
In the subsequent explanations of the simulation method, we will often refer to the contact network,
and to the consequence of adding or removing pairs of particles from it. 

\section{Simulational scheme}
\label{simulationDetails}
We now describe our simulation scheme; the scheme is identical for the case of spatial deformations or external
forcing. In the following we will refer to the contact forces or the velocities given by Eq.~(\ref{contactForces})
and Eq.~(\ref{eom}) respectively; the descriptions should be understood to equally apply 
using Eqs.~(\ref{strainContactForces}) and (\ref{strainVelocities}) for the case of spatial deformations. 

The next Subsection describes the basic integration loop of the method. 
We then dedicate Subsection (\ref{correctNetwork}) to the procedure which filters 
contacts carrying negative forces from the contact network. Subsection (\ref{correctionStep})
describes an algorithm which allows for the preservation of the accuracy of the contact network during long 
simulation runs. Subsection (\ref{jammed}) explains how jammed configurations can be generated using
our simulational method. 

\subsection{Basic integration loop}
The main idea is to evolve the system by integrating the equations of motion (\ref{eom}) in
time increments $\delta t$, until an occurrence of an event in which the contact network changes.
These instances are inferred by considering the following three conditions at all times:
\begin{itemize}
\item[($a$)]{The distance between the centers of particles that are in contact is (up to integration errors 
to be discussed in the following) the sum of their radii.}
\item[($b$)]{Particles that are not in contact do not overlap.}
\item[($c$)]{The contact forces $\ket{f}$ (given by Eq.~(\ref{contactForces})) between particles in contact are positive.}
\end{itemize}

The main simulation scheme is as follows:
\begin{itemize}
\item[($i$)]{
Given a contact network at some instance $t$ in which all contact forces are positive,
an integration step $\delta t$, and the particles' velocities 
$\ket{V}$, calculate the time $t_{\rm col}$ at which the next collision will occur.
The calculation of the next collision time is identical to the well-known calculation of 
the next collision time in event-driven hard sphere simulations, see for example~\cite{91AT}. 
However, the set of pairs of particles considered for a possible future collision
are only those which are {\bf not} already in contact.
}
\item[($ii$)]{
Compare $t_{\rm col}$ with $t + \delta t$; then
\begin{itemize}
\item[($ii.a$)]{
If $t_{\rm col} < t + \delta t$, integrate Eq.~(\ref{eom}) up to $t_{\rm col}$, and
add the colliding pair of particles to the contact 
network. This addition may render some of the other contact forces of the network negative. If 
indeed the collision results in the appearance of negative contact forces,
they are removed at this point by the \emph{rewiring procedure} described in Subsect.~\ref{correctNetwork}. 
}
\item[($ii.b$)]{
If $t_{\rm col} > t + \delta t$, integrate Eq.~(\ref{eom}) up to $t + \delta t$,
calculate the contact forces at the new state via Eq.~(\ref{contactForces}), and check whether any of the
contact forces has become negative. Remove the contacts with negative forces from the contact network using
the rewiring procedure described in Subsect.~\ref{correctNetwork}. 
}
\end{itemize}}
\item[($iii$)]{
Step $(ii)$ produces an updated contact network with strictly positive contact forces
and no overlapping particles. The new velocities $\ket{V}$ are then calculated
via Eq.~(\ref{eom}). Goto $(i)$.}
\end{itemize}

The integration step $\delta t$ in our simulations is set by requiring that the number of integration steps between
events (new contact formations or vanishing contact forces) is on average larger than 10. 
Satisfying this criterion requires gradually smaller integration steps for larger or denser systems. 
Moreover, in our simulations we bound the integration step by requiring $\delta t < 10^{-5}$. 
Detection of newly formed contacts
is acheived by maintaining data structures of neighbor-lists which are updated
using cell-subdivisions that allow for the calculation of the neighbor-lists in a running-time which is
linear in the system size \cite{91AT}.

In the following subsection we will describe and explain the rewiring procedure; in this procedure contact forces which are rendered negative,
either by the formation of a new contact, or by their natural evolution, are filtered out of the contact network. 

\subsection{The `rewiring' procedure - handling negative contact forces}
\label{correctNetwork}
As mentioned above, the creation of a new contact may lead to the appearance of negative forces in other pre-existing
contacts of the contact network. The determination of the new contact network after detection of negative contact forces
is carried out by repeating the following two procedures:
\begin{itemize}
\item[($i$)]{
First, we successively remove the contact with the most negative force from the contact network, 
and re-calculate the contact forces via Eq.~(\ref{contactForces}) after each such removal.
The identity of all the contacts removed from the contact network during this procedure is stored.
This removal procedure is carried out iteratively until the contact network is free of negative forces.
Then, the velocities are calculated using Eq.~(\ref{eom}). }
\item[($ii$)]{
With the velocities $\ket{V}$ calculated at the end of 
procedure $(i)$, we now consider the list of \emph{removed contacts} that we have constructed during
procedure $(i)$. For each removed contact, we calculate the pairwise velocity
\begin{equation}
v_{ij} \equiv \dot{r}_{ij} = \vec{n}_{ij}\cdot\left(\vec{V}_j - \vec{V}_i\right)\ ,
\end{equation}
where the pair $(i,j)$ no longer belongs to the contact network.
Note that since we are considering pairs of particles which are no longer 
in contact, constraint (\ref{constraint}) does not apply, and generally $v_{ij} \ne 0$.
For some of the pairs considered one may find that $v_{ij} < 0$ which means that
the particles $(i,j)$ which were a part of the contact network are now
{\bf approaching} each other after breaking the contact between them. 
This scenario violates the no-overlapping condition, since this pair 
is (idealy) just in contact (it was just removed from the contact network),
hence if this pair has $v_{ij} < 0$, it will instantly create an overlap.
We treat this scenario explicitly -- the pair with the most negative $v_{ij}$ is then re-connected
and added back to the contact network, and the contact forces are
then re-calculated. If there remain any negative contact forces $(f_{ij}< 0)$, or if 
pairs of particles {\bf that were removed} from the contact network are
still approaching each other $(v_{ij} < 0)$, return to ($i$). Otherwise, 
the final contact network has been found. }
\end{itemize}

The above procedure ends having in hand a contact network which satisfies conditions (a),(b) and (c) detailed in the beginning 
of this section. The uniqueness of the contact network obtained using the above procedure is demonstrated in Appendix~A.

\subsection{Maintaining correctness and error estimations}
\label{correctionStep}
The integration errors in distances between pairs
of particles that belong to the instantaneous contact network are of order $\delta t^2$.
This can be shown by returning to Eq.~(\ref{constraint}), and writing it for a pair $i,j$ as
\begin{equation}\label{foo16}
v_{ij} = \dot{r}_{ij} = (\vec{V}_j - \vec{V}_i) \cdot \vec{n}_{ij} = 0\ ,
\end{equation}
thus the velocity differences $\vec{V}_j - \vec{V}_i$ are orthogonal to the unit vectors 
of pairs of particles in contact $\vec{n}_{ij}$, see Fig.~\ref{errorFig}. 
Although the errors in the pairwise distances are very small, they can accumulate over
many integration steps. To prevent the accumulation of large errors in cases where
the lifetime of contacts is long, we utilize the following 
correction procedure periodically:
\begin{figure}[!ht]
\centering
\includegraphics[scale = 0.45]{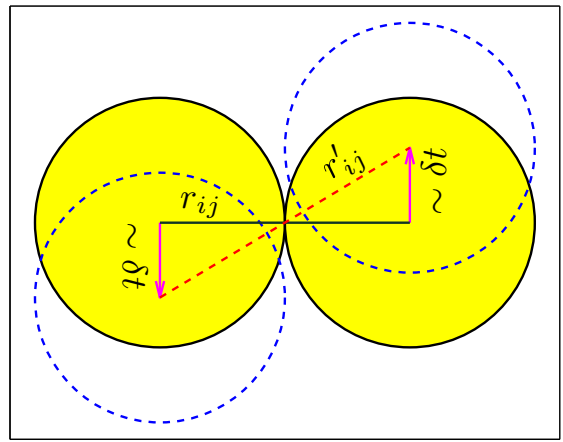}
\caption{\footnotesize The continuous (dashed) particles are depicted before (after) a displacement of order $\delta t$. 
The change in the distance between two particles in contact is of order
$r_{ij}' - r_{ij} \sim \delta t^2$ due to the orthogonality of the total velocity differences with respect
to the unit vector $\vec{n}_{ij}$, see Eq.~(\ref{foo16}).}
\label{errorFig}
\end{figure}
\begin{itemize}
\item[($i$)]{
For each pair $i,j$ in the contact network, calculate the pairwise errors $\delta r_{ij} = r_{ij} - (d_i + d_j)$,
where $d_i$ is the radius of the $i^{\rm th}$ particle. 
}
\item[($ii$)]{Denoting the pairwise errors as $\ket{\delta r}$, we 
calculate the correction displacement field
\begin{equation}\label{correction}
\ket{\delta R} = -{\cal S}^T{\cal N}^{-1}\ket{\delta r}\ .
\end{equation}
}
\item[($iii$)]{At this stage the displacement field $\ket{\delta R}$ is treated as a velocity field;
we then evolve the system in the direction determined by $\ket{\delta R}$, 
either until the entire displacement is achieved, or up to a collision induced by the evolution. 
Given displacements $\ket{\delta R}$
we calculate the next ballistic collision time $t_{\rm col}$ of pairs of particles that do not belong 
to the contact network, and record their identities. 
\begin{itemize}\item[$(iii.a)$]{
If $t_{\rm col} < 1$, update the coordinates via $\ket{R} \leftarrow  \ket{R} + t_{\rm col}\ket{\delta R}$,
and add the colliding pair of particles to the contact network. 
}
\item[$(iii.b)$]{
If $t_{\rm col} > 1$, 
update the coordinates via $\ket{R} \leftarrow  \ket{R} + \ket{\delta R}$.
}
\end{itemize}
}
\item[($iv$)]{If a new contact was formed in step $(iii)$, repeat steps $(i)-(iii)$ untill no new contacts are formed.}
\item[($v$)]{Calculate the contact forces, and filter out negative contact forces 
using the rewiring procedure described in Subsect.~\ref{correctNetwork}. 
}
\end{itemize}

This correction procedure eliminates the accumulated errors in the pairwise distance between particles which
are part of the contact network; the correction displacements $\ket{\delta R}$ can be thought of as the response of 
the system to an inhomogeneous expansion of the pairwise distances of the contacts within the contact network.
The frequency of application of the correction procedure can be tuned to achieve the desired accuracy of
the pairwise distances within the contact network.

\subsection{Generating jammed configurations}
\label{jammed}
Our simulation can be used to generate jammed configurations, under shear deformation
or under compression (or both). Before explaining how such configurations can be generated,
we first clarify the distinction between jammed and potentially jammed configurations. 

In a \emph{potentially jammed} configuration the ratio between the number of contacts and the number 
of degrees of freedom is such that the matrix ${\cal S}^T$ must have a kernel of dimension one. Thus there exists a 
non-trivial vector of contact forces $\ket{f_0} \ne 0$ that satisfies force balance, i.e. that solves the equation 
${\cal S}^T\ket{f_0} = 0$. However, a potentially jammed configuration may not be jammed,
since the vector $\ket{f_0}$ that solves ${\cal S}^T\ket{f_0} = 0$ may have 
negative components, which cannot represent contact forces between rigid, frictionless particles. 
Therefore, in a \emph{jammed} configuration there must exist a 
vector of {\bf positive} contact forces $\ket{f_0} > 0$,
that solves the equation ${\cal S}^T\ket{f_0} = 0$.

In order to generate such jammed configurations, one must be in the position to handle 
the problematic cases in which the system becomes potentially jammed. In these cases, 
the equation ${\cal S}^T\ket{f} = 0$ has a non-trivial solution, which means 
that ${\cal N} \equiv {\cal S}{\cal S}^T$ has a zero eigenvalue,
which in turn implies that Eq.~(\ref{contactForces}) cannot be solved,
thus the contact forces and the velocities cannot be calculated.
The detection of potentially jammed or jammed configurations requires the tracking of the coordination number
of the \emph{constrained subset} of the system. The constrained subset is the set of particles
which are in contact with {\bf at least} $d+1$ other particles from the constrained subset (where $d$ is the 
spatial dimensionality of the system considered). The determination of 
the constrained subset must be performed iteratively, since the removal of some of the 
particles from the constrained subset may exclude other particles connected to them
from the constrained subset as well. From a physical perspective, 
the monitoring of the constrained subset eliminates the problems of counting 
degrees of freedom associated with rattlers or particles which are just loosely connected
in the contact network. 

With the correct detection of the constrained subset in hand, we next define the \emph{constrained
coordination number} $\bar{z}$ as
\begin{equation}\label{foo06}
\bar{z} \equiv \frac{1}{\bar{N}}\left(2(d-1) +  \sum_{i \,\in\,{\mbox{\tiny{constrained subset}}}}\bar{z}_i\right)\ ,
\end{equation}
where $\bar{z}_i$ is the number of contacts between the $i^{\rm th}$ particle and
other particles $j$ which have $\bar{z}_j \ge d+1$,
and $\bar{N}$ is the number of particles $i$ with $\bar{z}_i \ge d+1$. 
The system is potentially jammed if $\bar{z} = 2d$.
Eq.~(\ref{foo06}) is obtained by comparing the number of contacts in the constrained
subset $\frac{1}{2}\sum_i \bar{z}_i$ to the number of degrees of freedom $\bar{N}d - d + 1$. 
The addition of one to the number of degrees of freedom assures that 
${\cal S}^T$ has a kernel of dimension one if the system is potentially jammed. 
Definition (\ref{foo06}) should be calculated using integer arithmetics,
since it is an exact indicator of isostaticity \cite{footnote1}.

Once a potentially jammed configuration has been detected, the non-zero vector $\ket{f_0}$ which solves 
${\cal S}^T\ket{f_0} = 0$ is calculated \cite{footnote3}. If all its components are positive,
the system is jammed. Otherwise, the rewiring procedure described in Subsect.~\ref{correctNetwork}
for filtering negative contact forces out of the contact network is employed,
assigning the solution $\ket{f_0}$ as the contact forces. 
Within the procedure, the constrained coordination number must
always be tracked; whenever the system is detected to be potentially jammed, 
the solution $\ket{f_0}$ should be re-calculated
and considered instead of the contact forces. 
The procedure ends with either $(i)$ some contacts being removed
such that the system is no longer potentially jammed, then the forces and velocities can be calculated as usual
usings Eqs.~(\ref{contactForces}) and (\ref{eom}) respectively, or $(ii)$ a contact network being 
formed with a solution $\ket{f_0}$ to ${\cal S}^T\ket{f_0} = 0$, having
only positive components, then a jammed configuration has been found.

\begin{figure}[!ht]
\centering
\includegraphics[scale = 0.27]{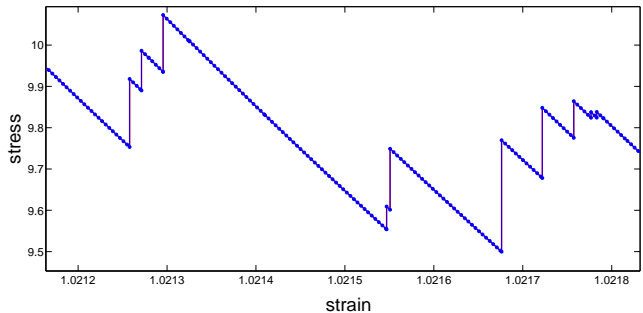}
\caption{\footnotesize Segment of a typical stress vs.~strain signal from our event-driven simulation of shear flow
in two dimensions at the volume fraction of $\phi = 0.83$. The stress relaxes during the time intervals in which the contact
network remains intact. The sudden upward jumps in stress signal the collision of particles.
Our simulation method enables an accurate study of the statistics of the collisions and their effect on the mechanical 
response of the system.}
\label{stressStrainFig}
\end{figure}

\section{Applications}
\label{results}
\subsection{Shear flow}
We simulate systems of discs in two dimensions, under simple shear flow \cite{12LDW}.
Our system consists of a binary mixture, where half of the discs are small and 
half are large. We add a uniform polydispersity of 3\% to avoid hexagonal patches of 
identical particles \cite{footnote2}. Our unit of length $\lambda$
is the mean diameter of the small particles. The large particle have a mean radius of 
$0.7\lambda$ and the small particles have a mean radius of $0.5\lambda$. 
The damping coefficient $\xi_0$ is set to unity, as is the strain-rate $\dot{\gamma} = 1$,
thus forces are measured in units of $\lambda\dot{\gamma}/\xi_0$.
The velocity field of the underlying fluid is $\vec{V}^{\rm aff} = \dot{\gamma}y\hat{x}$.
Simple shear flow is imposed homogeneously throughout the simulation cell using
Lees-Edwards boundary conditions \cite{91AT}.  The equation of motion (\ref{eom}) is integrated by
solving Eq.~(\ref{contactForces}) using a standard conjugate gradient algorithm \cite{92PTVF}.

In Fig.~(\ref{stressStrainFig}) we plot a small segment of a typical stress vs.~strain signal from our simulations at
a volume fraction of $\phi = 0.83$. The stress $\sigma$ is measured using information of the contact forces $f_{ij}$, and
the differences $\vec{R}_{ij}\equiv \vec{R}_j - \vec{R}_i$ via \cite{91AT}
\begin{equation}
\sigma = -\frac{1}{\Omega}\sum_{{\mbox{\tiny{contacts }}}ij}\frac{f_{ij}\left(\vec{R}_{ij}\cdot \hat{x}\right)\left(\vec{R}_{ij}\cdot\hat{y}\right)}{r_{ij}}
=  -\frac{\bra{f}{\cal S}\ket{V^{\rm aff}}}{\Omega}\ ,
\end{equation}
where $\Omega$ is the volume of the system. The signal displays a series of intervals in which the stress
relaxes, interrupted by abrupt increases in the stress, which are the consequence of collisions (new 
contacts are formed).

\begin{figure}[!ht]
\centering
\includegraphics[scale = 0.4]{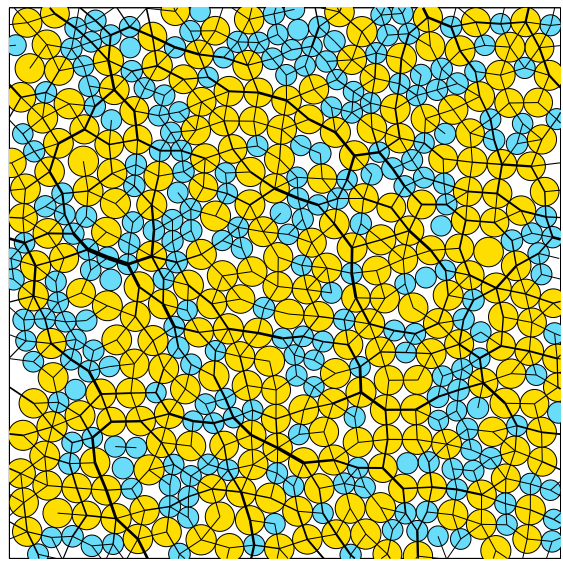}
\caption{\footnotesize A snapshot from our two-dimensional simulations of bi-disperse discs under shear flow.
The thickness of the lines connecting the centers of the particles represent the magnitude
of the contact forces between the particles. The colors distinguish between the two sizes of discs. }
\label{contactForcesFig}
\end{figure}

In Fig.~(\ref{contactForcesFig}) we display a realization of our sheared systems at $\phi=0.83$.
The thickness of the lines connecting the centers of neighboring particles are proportional 
to the contact force between the particles. 

\begin{figure}[!ht]
\centering
\includegraphics[scale = 0.45]{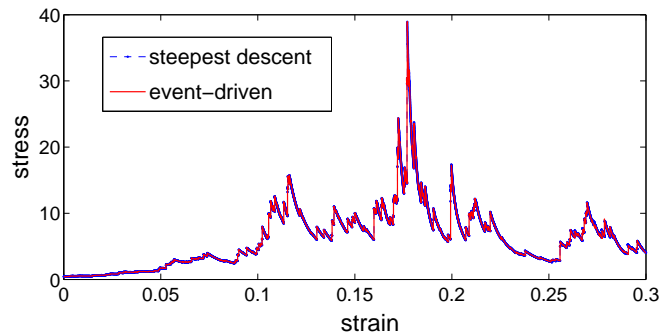}
\caption{\footnotesize Comparison between stress vs.~strain trajectories from our event-driven simulation
and from the steepest descent method. Starting from the same initial configurations, the trajectories remain
essentially identical for tens of percents of strain, vindicating our event-driven scheme.}
\label{constantPressureFig}
\end{figure}

\begin{figure}[!ht]
\centering
\includegraphics[scale = 0.43]{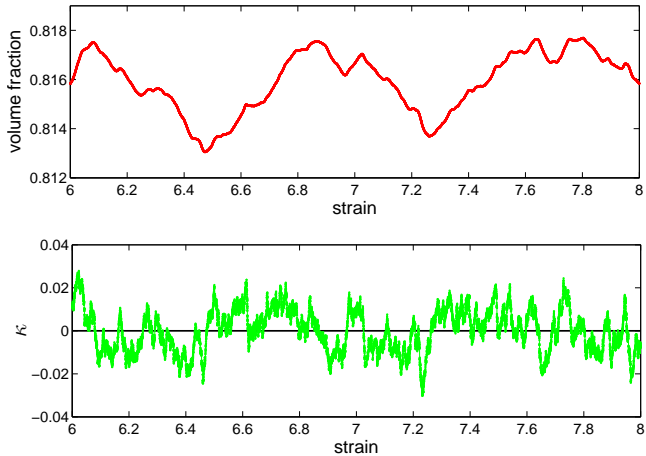}
\caption{\footnotesize Segments of typical signals from simulations of shear flow of bi-disperse discs
under a constant imposed pressure of $p=10$ (in reduced units). 
Upper panel: volume fraction $\phi$ vs.~strain. Lower panel: 
the ratio of the rate of isotropic compression 
and the rate of simple shear, $\kappa$, vs.~strain. Steady-state implies that $\kappa$ must fluctuate around
zero, as on the average the volume fraction is stationary for a given pressure.}
\label{compare}
\end{figure}

\subsubsection{Comparison with steepest descent simulations}
An alternative to integrating our equation of motion, Eq.~(\ref{eom}), one can apply small strain-steps
followed by the minimization of a soft, repulsive interaction potential
by means of steepest descent, i.e. according to the first order equation of motion
\begin{equation}
\dot{\vec{R}}_k = -\xi_0\nabla_k U\ ,
\end{equation}
where a possible choice of the potential energy is $U = \sum_{i,j>i}\phi(r_{ij})$, and 
\begin{equation}\label{potential}
\phi(r_{ij}) = \left\{
\begin{array}{ccc}
\varepsilon\left(1-\frac{r_{ij}}{d_i + d_j}\right)^2 &,&  r_{ij} < d_i + d_j \\
0 &, & r_{ij} > d_i + d_j \\
\end{array}\right. \ ,
\end{equation}
where $\varepsilon$ is an energy scale and $d_i$ is the radius of the $i^{\rm th}$ particle.
The equivalence between the two approaches stems from $(i)$ the identical dissipation mechanism, which 
only acts along the the nonaffine part of the motion, and $(ii)$ the total evolution
that keeps the potential energy at zero since the system is studied in the floppy regime, where
it is always possible to re-arrange the particles to a state of zero potential energy.
This is analogous in our simulation to the no-overlap condition Eq.~(\ref{constraint}) which is always satisfied. 
To validate our simulational scheme, we compare the results of our scheme with the steepest descent (SD)
simulations. In the SD simulation, we construct a tentative contact network by 
considering pairs of particles that satisfy $r_{ij} = d_i + d_j + \delta r$ with $\delta r = 10^{-5}$. 
This network is constructed only for the contact force calculation from which the stress 
is deduced. We choose the strain increment for the SD simulations to be $\delta\gamma = 10^{-6}$ and 
choose the stopping condition for the minimization to be $\max\left( |\nabla_i U \right|) < 10^{-9}\varepsilon/\lambda$. 
The integration step for the steepest descent minimizations was set to be $\delta t = 0.1$. 
In Fig.~\ref{compare} we plot the trajectories of the stress vs.~strain for both simulation schemes,
for systems of $N=100$ particles at the volume fraction $\phi = 0.82$.
The agreement between the two signals vindicates our event-driven scheme. We note that
for strain increments of $\delta\gamma = 10^{-6}$, with which
the events of contact formations can be clearly singled out in the SD simulations, 
the running time is extremely long, hence the choice of small systems for comparison with our method. 
It took over 30 hours on a conventional workstation to produce the 
data of Fig.~\ref{compare} using the SD method,
whereas it takes just a few minutes with our event driven code on the same machine.

\subsubsection{Simulations of shear-flow under fixed pressure}
We also simulated the same system described above, but with imposing 
a fixed pressure, allowing for the volume to fluctuate (see Subsection~\ref{constantPressure} for
details about the method).
The same simulation scheme described in Sect.~(\ref{simulationDetails}) can be used,
with the additional step of calculating $\kappa$ via Eq.~(\ref{kappa})
in every calculation of the contact forces, Eq.~(\ref{constantPressureForces}).
Then, $\kappa$ is used in Eq.~(\ref{constantPressureAffineVelocities}) to determine the affine velocities, which, 
by construction, produce the chosen imposed pressure $p$ under shear flow.
In Fig.~(\ref{constantPressureFig}) we present typical signals of the
volume fraction and the parameter $\kappa$ which represents the ratio of the rate of isotropic 
compression and the rate of simple shear.

\begin{figure}[!ht]
\centering
\includegraphics[scale = 0.75]{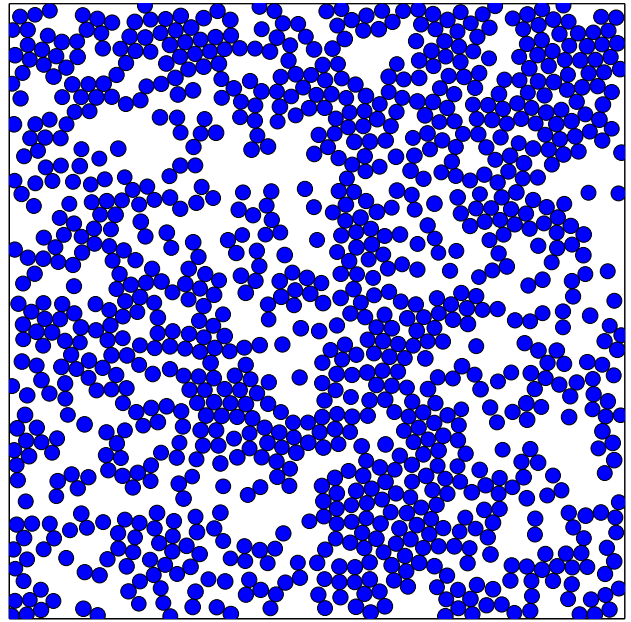}
\caption{\footnotesize Snapshot from event driven simulations of overdamped self-propelled discs in two dimensions.
The packing fraction is $\phi = 0.5$ and the persistence duration of the propelling forces is $10\tau$.}
\label{propelledFig}
\end{figure}

\subsection{Self-propelled particles}
We employed our event-driven method to simulate overdamped, self-propelled discs in two dimensions. 
The diameters of the discs were chosen with a uniform polydispersity of 3\% \cite{footnote2},
with a mean of unity. For each particle $i$ a random force $\vec{F}^{\rm ext}_i$ is chosen and 
imposed a persistence duration $\tau_i$, after which a new duration $\tau_i$ and 
and new force $\vec{F}^{\rm ext}_i$ are chosen and imposed. 
The propelling force was randomly chosen from a uniform distribution, such that 
$f^{\rm ext} \equiv \sqrt{\braket{F^{\rm ext}}{F^{\rm ext}}}/N = 1$. 
The persistence duration of the propelling force on each particle was uniformly chosen from
the interval [$0\tau$,10$\tau$], where 
$\tau = \lambda/(f^{\rm ext}\xi_0)$, $\lambda$ is the mean particle diameter and 
$\xi_0$ is the proportionality constant that relates velocities to forces in our overdamped 
dynamics (see Eq.~(\ref{dragForces})). Fig.~(\ref{propelledFig}) displays a snapshot from 
our simulations of self-propelled discs.

\subsection{Composite-particles}
\begin{figure}[!ht]
\centering
\includegraphics[scale = 0.44]{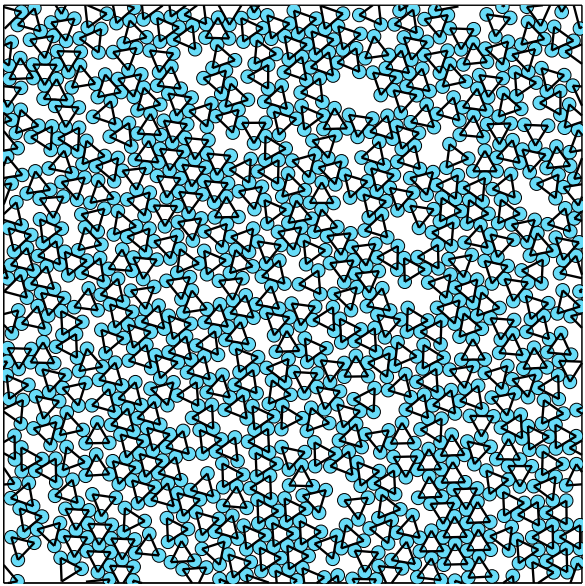}
\caption{\footnotesize Snapshot from event driven simulations of composite particles under shear flow in two dimensions.
The black lines connecting triples of particles represent the permenent contacts in each composite particle, which behaves
as a rigid body.}
\label{compositeFig}
\end{figure}

We end this section with a demonstration of how our method can be used to easily simulate
composite-particles made of a number of beads attached together to form rigid bodies, under external
driving. The method is employed as explained in the previous sections, except that pre-assigned contacts 
between particles belonging to the same composite particle are never broken. 
Clearly this requires us to allow for negative contact forces between the particles which belong to the same composite-particle. 
Other than this classification of contacts as permanent (allowed to bear negative forces) 
and temporary (can only carry positive forces), the simulation scheme is identical to that described above. 
Fig.~(\ref{compositeFig}) displays a snapshot from our simulations of composite particles under shear-flow.

\section{Discussion}
\label{conclusions}
We have introduced an event-driven simulation scheme for overdamped hard spheres
under external driving forces. The simulation is based on 
an equation of motion which is exactly derivable from the model
assumptions. The simulation maintains 
the conditions of no overlaps and strictly positive contact forces between 
the rigid particles at all times. We have introduced a procedure that 
allows for the correction of the integration errors in the pairwise distances between 
particles that are part of the instantaneous contact network.
We have also provided a perscription for generating jammed 
configurations employing our method,
under compression or shear.

We have demonstrated that our method produces essentially identical
results to steepest-descent minimization methods, though at greatly reduced 
computational cost as well as increased robustness and accuracy of the contact network information.  		
Indeed, the main virtue of our event-driven method is the control over 
the identity of the contact network that allows for careful 
study of the statistics of contact formation and breakage. 
In our approach there are no uncertainties nor arbitrary thresholding
associated with the definition of which particles should be considered as contacts. 
This advantage becomes increasingly important  closer to  jamming where contacts rearrange very rapidly,
and where many particles are very close but actually not in contact.

A simple extension of the formalism presented here allows us to simulate systems under 
shear flow and {\bf fixed} imposed pressure, without introducing any free parameters.
This is similar to constraint methods used
in conventional molecular dynamics simulations to maintain a constant pressure or temperature \cite{91AT}.
This extension of our method may be useful in simulating shear flow of frictionless hard spheres
close to the jamming point \cite{07OT,11OT,09HB,12LDW}: in constant volume simulations, 
fluctuations of the critical packing fraction $\phi_c$ can lead to jamming and complete arrest of the flow. 
When fixing the pressure instead of the volume, structural fluctuations which would lead to 
jamming in fixed-volume simulations, merely lead to the dilatancy of the system as it expands under 
the increasing internal pressure, and avoiding jamming by doing so. 

Finally, we have demonstrated a number of possible additional applications of our method. 
We note here that the addition of boundaries and various container shapes is straightforward
within our framework, since they can be treated on equal grounds as the contacts between
the particles.

\appendix
\label{appendix}
\section{ Uniqueness of the contact network upon new contact formation}
The algorithm described in Subsect.~\ref{correctNetwork} begins with the full set of contacts,
including a newly formed contact. 
We denote this full set of contacts by $\Gamma$. Upon termination, the algorithm produces 
a subset of contacts, denoted here by $A$, which satisfies
the two following conditions:
\begin{itemize}\item[$(i)$]{
all of the contact forces calculated on the subset $A$ of contacts are positive, and }
\item[$(ii)$]{
all of the pairs of particles which were \emph{just removed} from the contact network,
are moving away from each other.}
\end{itemize}
In this subsection we prove that the contact network found by our algorithm is unique. 
We first re-iterate that the operators
${\cal S}$ (Eq.~(\ref{esDefinition}) and ${\cal N} \equiv {\cal S}{\cal S}^T$  
are generally defined on a given set of contacts.
When a new contact is formed, the set of pairs on which these 
operators are defined grows by one pair -- the new contact. 
However, as mentioned above, the formation of a new contact may
render other contact forces negative, in which case those pairs with negative contact forces
are removed from the contact network.
Removing contacts from the contact network re-defines the operators 
${\cal S}$ and ${\cal N}$, since the space of contacts has changed. 
Given the subset of contacts $A \subseteq \Gamma$, we
denote the corresponding operators defined on this set as ${\cal S}_A$ and ${\cal N}_A$. 
As opposed to the space of contacts, the space of particles never changes;
one can, in principle, calculate the full vector of velocities given any subset of the full contact
space. For instance, the velocities calculated using the subset $A$ are (adopting notations
and formulae from Subsect.~(\ref{spatialDeformations}))
\begin{equation}\label{foo05}
\ket{V_A} = \ket{V^{\rm aff}} - {\cal S}_A^T{\cal N}_A^{-1}{\cal S}_A\ket{V^{\rm aff}}\ .
\end{equation}

We now present the general relation between contact forces and velocities;
starting from Eq.~(\ref{moo01}), and setting $\xi_0 = 1$
for the sake of brevity, we consider the full contact network, then
\begin{equation}
{\cal S}^T\ket{f} = \ket{V} - \ket{V^{\rm aff}}\ .
\end{equation}
Operating with ${\cal S}$ on both sides of the above equation, 
\begin{equation}
{\cal S}{\cal S}^T\ket{f} = {\cal N}\ket{f} = {\cal S}\ket{V} - {\cal S}\ket{V^{\rm aff}}\ .
\end{equation}
Now, inverting to solve for the contact forces
\begin{equation}\label{foo03}
\ket{f} = {\cal N}^{-1}{\cal S}\ket{V} - {\cal N}^{-1}{\cal S}\ket{V^{\rm aff}}\ .
\end{equation}
Given any velocity vector $\ket{V}$, the resulting contact forces can be obtained via Eq.~(\ref{foo03}). 
This equation is more general than Eq.~(\ref{contactForces}), that can be obtained from 
the above relation by demanding that ${\cal S}\ket{V} = 0$, which is the 
condition for maintaining an intact contact network. In our discussion we 
relax this condition for a subset of the contact network, since removing 
contacts will result in {\bf non-zero} pairwise velocities. 

We next note that a contact force which is identically zero has no effect on the 
dynamics. This can be understood by re-examining Eq.~(\ref{moo01});
a zero contact force does not participate in the vector sum of the contact forces on each particle, 
which, in turn, is equal to the nonaffine part of the velocities $\ket{V} - \ket{V^{\rm aff}}$. This allows us to carry out the
entire discussion in the full space of contacts $\Gamma$; 
the vector of contact forces in the full space of contacts, obtained with the velocities 
calculated on the subset $A$, is
\begin{equation}
\ket{f_A} = {\cal N}^{-1}{\cal S}\ket{V_A} - {\cal N}^{-1}{\cal S}\ket{V^{\rm aff}}
\end{equation}
Note that the dimension of $\ket{f_A}$ is the size of the full set of contacts $|\Gamma|$.
Labeling contact components by $\alpha \in \Gamma$, and denoting by $\ket{\delta_\alpha}$ a vector which is 
unity in the $\alpha$ component and zero otherwise, the condition of positive contact forces now reads
\begin{equation}\label{foo10}
\begin{array}{ccc}
\braket{\delta_\alpha}{f_A} > 0&\mbox{if}& \alpha \in A\ , \\
\braket{\delta_\alpha}{f_A} = 0&\mbox{if}& \alpha \in \Gamma\setminus A\ .
\end{array}
\end{equation}
The vector of pairwise velocities induced by the particles' velocities $\ket{V_A}$ Eq.~(\ref{foo05}) is  
\begin{equation}
\ket{v_A} = {\cal S}\ket{V_A}\ .
\end{equation}
The condition according to which pairs of particles that were just removed from 
the contact network should be moving apart can be written in our framework as
\begin{equation}\label{foo11}
\begin{array}{ccc}
\braket{\delta_\alpha}{v_A} = 0&\mbox{if}& \alpha \in A\ , \\
\braket{\delta_\alpha}{v_A} > 0&\mbox{if}& \alpha \in \Gamma\setminus A\ .
\end{array}
\end{equation}

We now assume that there exists another {\bf different} subset
$B\subseteq \Gamma, B \ne A$, and an associated velocity vector 
$\ket{V_B} = \ket{V^{\rm aff}} - {\cal S}_B^T{\cal N}_B^{-1}{\cal S}_B\ket{V^{\rm aff}}$,
the contact forces vector $\ket{f_B}$ and the pairwise velocities $\ket{v_B}$, 
which also satisfy (\ref{foo10}) and~(\ref{foo11}) for the subset $B$. 
We consider the contraction
\begin{equation}\label{foo04}
\bra{v_A - v_B}{\cal N}^{-1}\ket{v_A - v_B} > 0\ ,
\end{equation}
which is positive since ${\cal N} \equiv {\cal S}{\cal S}^T$ is positive definite. 
Combining $\ket{v} = {\cal S}\ket{V}$ with Eq.~(\ref{foo03}) we have
\begin{equation}
\ket{f_A - f_B} = {\cal N}^{-1}{\cal S}\ket{V_A} - {\cal N}^{-1}{\cal S}\ket{V_B}
= {\cal N}^{-1}\ket{v_A - v_B}\ .
\end{equation}
Inserting this back into Eq.~(\ref{foo04}), we find
\begin{eqnarray}
\bra{v_A - v_B}{\cal N}^{-1}\ket{v_A - v_B}  =  \braket{v_A - v_B}{f_A - f_B} & & \nonumber \\
=  \braket{v_A}{f_A} + \braket{v_B}{f_B} - \braket{v_A}{f_B} - \braket{v_B}{f_A} & &\ .
\end{eqnarray}
From the conditions in Eq.~(\ref{foo10}) and Eq.~(\ref{foo11}), both 
$\braket{v_A}{f_A} = 0$ and $\braket{v_B}{f_B} = 0$.  The inequality (\ref{foo04})
now reads
\begin{equation}
0 < \bra{v_A - v_B}{\cal N}^{-1}\ket{v_A - v_B} =  - \braket{v_A}{f_B} - \braket{v_B}{f_A} < 0\ ,
\end{equation}
which is a contradition, since the cross terms
$\braket{v_A}{f_B}>0$ and $\braket{v_B}{f_A}>0$, as all components
of $v_A,v_B,f_A,f_B$ are non-negative by construction.
Hence, there cannot exist another distinct subset $B\ne A$ which satisfies Eqs.~(\ref{foo10}) and~(\ref{foo11}),
and therefore the subset $A$ is unique.

\subsection*{Generalization for the case of fixed pressure}
For the sake of brevity, we again set $\xi_0=1$ and also $\dot{\gamma} = 1$.
In the constant pressure formalism (Subsect.~\ref{constantPressure}), the affine
velocities satisfy
\begin{equation}
{\cal S}\ket{V^{\rm aff}} = -\kappa \ket{r} + \ket{\sFrac{r_xr_y}{r}}\ .
\end{equation}
Similarly to Eq.~(\ref{foo03}) the general relation between contact forces $\ket{f}$ and pairwise velocities $\ket{v}$ is
\begin{equation}\label{moo06}
\ket{f} = {\cal N}^{-1}\ket{v} + \kappa{\cal N}^{-1}\ket{r} - {\cal N}^{-1}\ket{\sFrac{r_xr_y}{r}}\ .
\end{equation}
Since this analysis includes the total set of contacts upon collision, 
we must redefine $\kappa$ of Eq.~(\ref{kappa})
in accordance with (\ref{moo06}) by requiring $p = \frac{\braket{f}{r}}{\Omega d}$, then
\begin{equation}\label{moo04}
\kappa = \frac{p\Omega d + \bra{r}{\cal N}^{-1}\ket{\sFrac{r_xr_y}{r}}
- \bra{r}{\cal N}^{-1}\ket{v}}{\bra{r}{\cal N}^{-1}\ket{r}}\ .
\end{equation}
We now assume that there exists $(\ket{v_A},\ket{f_A})$ and
$(\ket{v_B},\ket{f_B})$ such that 
\begin{eqnarray}
\braket{v_A}{f_A} & = & 0 \ , \nonumber \\
\braket{v_B}{f_B} & = & 0\ , \nonumber \\
\braket{v_A}{f_B} & > & 0 \ , \nonumber \\
\braket{v_B}{f_A} & > & 0 \ . \nonumber
\end{eqnarray}
Operating with ${\cal N}$ on Eq.~(\ref{moo06}), one finds
\begin{equation}\label{moo05}
{\cal N}\ket{f_B - f_A} = \ket{v_B - v_A} + (\kappa_B - \kappa_A)\ket{r}\ .
\end{equation}
Now, since ${\cal N}$ is positive definite
\begin{eqnarray}
\!\!\!\!\!\!\!\!0 & < &\!\! \bra{f_B - f_A}{\cal N}\ket{f_B - f_A} \nonumber \\
\!\!\!\!\!\!\!\!&= &\!\! \braket{f_B - f_A}{v_B - v_A} + (\kappa_B - \kappa_A) \braket{f_B\! -\! f_A}{r}.
\end{eqnarray}
Since $\braket{f_B}{r} = \braket{f_A}{r} = p\Omega d$ then
$\braket{f_B - f_A}{r} = 0$ and the inequality (\ref{moo04}) becomes
\begin{eqnarray}
0 & < & \braket{v_B}{f_B} - \braket{v_B}{f_A} - \braket{v_A}{f_B} + 
\braket{v_A}{f_A} \nonumber \\
& = & - \braket{v_B}{f_A} - \braket{v_A}{f_B} \ ,
\end{eqnarray}
which is a contradiction, since $\braket{v_B}{f_A}$ and $\braket{v_A}{f_B}$ are
strictly positive. This concludes the proof that there is only one subset which 
satisfies the constraints of positive forces and no overlaps, in the constant pressure framework.



\begin{thebibliography}{99}

\bibitem{86HM}
J.~P.~Hansen and I.~R.~McDonald, \emph{Theory of Simple Liquids} (Academic Press, London, 1986).

\bibitem{62AW}
B.~J.~Alder and T.~E.~Wainwright, Phys.~Rev.~{\bf 127}, 359 (1962).

\bibitem{00TS}
S.~Torquato and F.~H.~Stillinger, Rev.~Mod.~Phys.~{\bf 82}, 2633 (2010).

\bibitem{00CR}
G.~Combe and J.-N.~Roux, Phys.~Rev.~Lett.~{\bf  85}, 3628 (2000).

\bibitem{12IBS}
A.~Ikeda, L.~Berthier, and P.~Sollich, Phys. Rev. Lett.~{\bf 109} 018301 (2012).

\bibitem{07BW}
C.~Brito and M.~Wyart, J.~Stat.~Mech.~(2007) L08003; J.~Chem.~Phys.~{\bf 131}, 024504 (2009).

\bibitem{82Leu}
E.~Leutheusser, J.~Phys.~C {\bf 15}, 2801 (1982).

\bibitem{07OT}
P.~Olsson and S.~Teitel, Phys.~Rev.~Lett.~{\bf 99}, 178001 (2007).

\bibitem{11OT}
P.~Olsson and S.~Teitel, Phys. Rev. E, {\bf 83} 030302 (2011).

\bibitem{09HB}
C.~Heussinger and J.-L.~Barrat, Phys. Rev. Lett. {\bf 102}, 218303 (2009).

\bibitem{12LDW}
E.~Lerner, G.~D{\"u}ring, and M.~Wyart, Proc.~Natl.~Acad.~Sci, {\bf 109} 4798 (2012).

\bibitem{00Roux}
J.~-N.~Roux, Phys.~Rev.~E {\bf 61}, 6802 (2000).

\bibitem{91AT}
M.~P.~Allen and D.~J.~Tildesley, \emph{Computer Simulations of Liquids}
(Oxford Univ.~Press, New York, 1991). 



\bibitem{12LDWb}
E.~Lerner, G.~D{\"u}ring, and M.~Wyart, Europhys.~Lett.~{\bf 99} 58003 (2012).

\bibitem{09RR}
F.~Radjai and V.~Richefeu, Mech.~Mater.~{\bf 41}, 715 (2009).

\bibitem{91LSP}
B.~D.~Lubachevsky, F.~H.~Stillinger, and E.~N.~Pinson, J.~Stat.~Phys.~{\bf 64}, 501 (1991).

\bibitem{03OSLN}
C.~S.~O'Hern, L.~E.~Silbert, A.~J.~Liu, and S.~R.~Nagel, Phys.~Rev.~E {\bf 68}, 011306 (2003).

\bibitem{05Wyart}
M. Wyart, Annales de Physiques Fr., Chapter 7, {\bf 30}, 1, 2005. 

\bibitem{95Dur}
D.~J.~Durian, Phys.~Rev.~Lett.~{\bf 75}, 4780 (1995).

\bibitem{97Dur}
D.~J.~Durian, Phys.~Rev.~E {\bf 55}, 1739 (1997).

\bibitem{09Hat}
T.~Hatano, Phys. Rev. E {\bf 79} 050301 (2009)({\bf R}).

\bibitem{footnote1}
This indicator is exact if there are no over-constraints regions in the sample. This is the case for polydisperse particles
where crystalline over-coordinated regions cannot appear.

\bibitem{footnote3}
As $\ket{f_0}$ is the eigenvector associated with the zero eigenvalue of ${\cal N}$, it
can be calculated via diagonalization, for instance. 

\bibitem{footnote2}
Such patches can be hyperstatic, i.e.~over-constrained.
Physically these patches move a solid blocks, until they break. 
In our approach these patches are difficult to deal with (although it can be done),
as they lead to zero modes in the spectrum of ${\cal N}$, which
renders Eq.~(\ref{contactForces}) difficult to solve numerically.
We avoid this problem by adding polydispersity.


\bibitem{92PTVF}
W.~H.~Press, S.~A.~Teukolsky, W.~T.~Vetterling, and B.~P. Flannery, 
\emph{Numerical Recipes in C---The Art of Scientific Computing}, 2nd ed.~(Cambridge Univ. Press, Cambridge, 1994).


\end{thebibliography}
\end{document}